\documentclass[pra,twocolumn,showpacs,floatfix,showkeys,longbibliography]{revtex4-1}
\usepackage{times,amsmath,amssymb,amstext,latexsym,float,graphicx,color,ulem}
\usepackage{hyperref}
\hypersetup{colorlinks=true, citecolor=blue, urlcolor=blue, linkcolor=blue}

\begin{document}

\title{Toroidal Dipolar Supersolid with a Rotating Weak Link}

\author{M.~Nilsson Tengstrand}
\author{P.~Stürmer}
\email{philipp.sturmer@matfys.lth.se}
\author{J.~Ribbing}
\author{S.M.~Reimann}
\affiliation{Division of Mathematical Physics and NanoLund, Lund University, Box 118, SE-221 00 Lund, Sweden }

\date{\today}

\begin{abstract}
Ring-shaped superfluids with weak links provide a perfect environment for studying persistent currents and dynamic stirring protocols. Here, we investigate the effects of a weak-link system on dipolar supersolids. By calculating the ground state energy at fixed angular momenta, we find that metastable persistent currents may exist in the supersolid phase near the superfluid transition point. When stirring the weak link rapidly enough, we show that vortices can enter the supersolid. These vortex entries cause phase slips, emitting solitonic excitations that interfere with the crystalline structure of the supersolid, leading to a continuous melting and recrystallization of the droplets. Finally, we examine the release of vortex-carrying supersolids from the trap, observing that the released density exhibits a discrete structure associated with the density modulation and a central hole resulting from the vortex core.
\end{abstract}

\maketitle

\section{Introduction}
A many-body quantum state with off-diagonal long-range order exhibits dissipationless flow, known as superfluidity~\cite{ref:Gross1957,ref:Yang1962}. This leads to remarkable phenomena, such as quantized vortices and persistent currents, observed in electron systems due to Bose-condensed Cooper pairs, leading to SQUIDS and their applications in closed superconducting loops~\cite{ref:Jaklevic1964,ref:Sato2012}. Similarly, persistent currents occur in neutral systems like the early experiments in superfluid liquid helium~\cite{ref:allum1977,ref:Avenel1985,ref:hoskinson2006Helium} or Bose-condensed ultracold alkali atoms~\cite{ref:Legget1999,ref:mueller2002,ref:Ryu2007,ref:Moulder2012,ref:Beattie2013}. These persistent currents are metastable, have integer values of angular momentum per particle in units of $\hbar$ ~\cite{ref:Bloch1973}, and are insensitive to perturbations and disorder. Including a small potential barrier, a so-called `weak link', allows for states of different integer angular momentum to mix, effectively creating a two-level system~\cite{ref:Anderson2003,ref:Solenov2010a,ref:Solenov2010b,ref:Ryu2013,Cominotti2014,ref:Aghamalyan2015,ref:Amico2021,ref:Amico2022}. Stirring the weak link dynamically~\cite{ref:Wright2013a,ref:Wright2013b,ref:Piazza2013,ref:Ramanathan2011,ref:Mathey2014} can induce hysteresis between different persistent current states~\cite{ref:eckel2014}. As with any hysteresis process, the different states are well-separated by a potential barrier. This potential barrier, however, can be overcome with solitonic excitations which provide a steady-state solution and allow for non-quantized persistent currents~\cite{ref:Munoz2015}.\newline
If the many-body quantum state exhibits off-diagonal and diagonal long-range order simultaneously it results in the counter-intuitive state of supersolidity~\cite{ref:Chester1970,ref:Legget1970}, where superfluidity and crystalline structure coexist. While long sought after in $^4$He~\cite{ref:Kim2004,ref:Kim2012,ref:Balibar2010,ref:Boninsegni2012,ref:Kim2012,ref:Chan2013}, supersolidity has been observed in spin-orbit or light-coupled condensates~\cite{ref:Henkel2010,ref:lin2011,ref:li2016,ref:Leonard2017}, where, however, it occurs due to external factors and not due to the spontaneous symmetry breaking of both translational and gauge symmetry. The spontaneous breaking of these symmetries can be induced in ultracold dipolar Bose gases of dysprosium ($^{164}$Dy) and erbium ($^{164}$Er)~\cite{ref:Tanzi2019,ref:Bottcher2019,ref:Chomaz2019}, following the stabilization of a dipolar system against collapse when reducing its scattering length~\cite{ref:Kadau2016}. The stabilization is due to a beyond-mean-field correction term that is usually negligible and occurs due to the interplay of different density dependencies of the mean-field and beyond-mean-field terms~\cite{ref:Lee1957,Lima2011,Lima2012,ref:Petrov2015,Wachtler2016}. Only when tuning the system's scattering length to particular values via Feshbach resonances do these quantum fluctuations have a significant effect. The increase of the effective dipolar interaction strength $\varepsilon_{\mathrm{dd}}=a_{\mathrm{dd}}/a$, where $a_\mathrm{dd}$ is the dipolar length and $a$ the $s$-wave scattering length, beyond a critical value accompanies a softening of the roton mode in the exctiation spectrum~\cite{ref:Pomeau1994,ref:Saccani2012,ref:Macia2012,ref:Roccuzzo2019,ref:Natale2019}. The finite momentum value associated with the roton minimum relates to the periodicity of the crystalline structure within the supersolid or the isolated droplets~\cite{ref:Sanots2003}.\newline
\begin{figure}[H]
\centering
\includegraphics[width = 0.4\columnwidth]{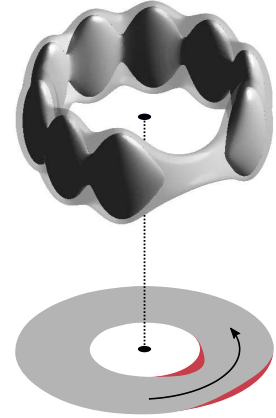}
\caption{({\it Color online}) Density isosurfaces for a toroidal supersolid with $\varepsilon_\mathrm{dd} = 1.366$ and our wide link system, where the transparent and solid isosurfaces are taken at density values $n=7.5\times 10^{13}$ cm$^{-3}$ and $n=2\times 10^{14}$ cm$^{-3}$, respectively. The grey-shaded area is the contour of the toroidal trap with a weak link, and the red-shaded area is the difference from a purely toroidal trap.}
\label{fig:schematic}
\end{figure}
In previous work, we found a hysteretic cycle in the angular momentum states in a toroidal supersolid of dipolar dysprosium atoms and demonstrated the existence of persistent currents in these systems~\cite{ref:NilssonTengstrand2021}. We now turn to a similar system, however, with a weak link (see Fig.~\ref{fig:schematic}). The weak link causes density modulations to form around it, close to the regime where the roton softens, similar to a dark solitary wave in a dipolar superfluid~\cite{ref:Edmonds2016}. This weak link creates an avoided level-crossing of angular momentum states, depending on the relative dipolar strength and the weak link strength. When rotating the weak link, the droplets move along while the superfluid stays stationary. However, as with purely contact-interacting superfluids, there is a local flow within the weak link~\cite{ref:Dubessy2012,ref:Woo2012,ref:Mathey2014}. If the flow velocity in this area approaches the local speed of sound it becomes energetically favorable for a supercurrent to exist. With the emergence of the persistent current, a phase slip and a corresponding vortex entry occur. We differentiate between two different regimes. In the well-established superfluid phonon-dominated regime, a solitonic wave-like excitation accompanies the phase slip. The solitonic excitation decays over time as it creates phonon excitations in the system. Further, the exact type of solitonic excitation depends on the chemical potential. In a regime where the roton mode softens, however, a roton excitation accompanies the emerging dark solitary wave. This roton leads to a density modulation and a still strongly connected supersolid. By bringing the system past the critical point into the supersolid phase, the emitted roton and dark solitary excitation lead to complicated dynamics between the droplets, resulting in a roton-mitigated melting and recrystallization of neighboring crystal sites, reminiscent of the Lindemann melting criterion~\cite{ref:Lindemann1910,ref:Khrapak2020}.
\section{Model and Methods}

We model our dipolar condensate using a nonlocal extended Gross-Pitaevskii equation
\begin{equation}\label{egpe}
i \hbar \partial_t \psi(\mathbf{r}, t) = H_\mathrm{GP}\psi(\mathbf{r}, t),
\end{equation}
\noindent where
\begin{equation}\label{gph}
\begin{aligned}
H_\mathrm{GP} =  &-\frac{\hbar^2 \nabla^2}{2m} + V_\mathrm{trap}(\mathbf{r}) + g|\psi(\mathbf{r},t)|^2  \\
&+ \int d \mathbf{r}' V_\mathrm{dd}(\mathbf{r}-\mathbf{r}') |\psi(\mathbf{r}', t)|^2 + \gamma |\psi(\mathbf{r},t)|^3.
\end{aligned}
\end{equation}
\noindent Here the order parameter $\psi$ is normalized to the total number of particles $N$, $m$ is the mass of a single particle, and $g=4\pi \hbar^2 a/m$ is the contact interaction coupling constant with the $s$-wave scattering length $a$. The nonlocal dipolar potential is $V_\mathrm{dd}(\mathbf{r}) = \frac{\mu_0 \mu^2}{4\pi} \frac{1-\cos^2\beta}{|\mathbf{r}|^3}$, where $\mu_0$ is the permeability in a vacuum, $\mu$ the magnetic moment, and $\beta$ the angle between $\mathbf{r}$ and the dipole alignment direction, which is taken to be along the $z$-axis. The final term is the beyond-mean-field term in the local density approximation \cite{Lima2011, Lima2012, Wachtler2016}, where $\gamma = \frac{128 \sqrt{\pi}}{3} \frac{\hbar^2 a^{5/2}}{m}(1 + \frac{3}{2}\varepsilon_\mathrm{dd}^2)$, and $a_\mathrm{dd} = m\mu_0 \mu^2 / (12\pi\hbar^2)$. The ring-shaped trap with a weak link is modeled according to
\begin{equation}
V_\mathrm{trap}(\mathbf{r}) = \frac{m\omega^2(\theta)}{2}\left[(\rho-\rho_0)^2 + \lambda^2 z^2\right],
\end{equation}
\noindent where the dependence on the azimuthal angle $\theta$ of $\omega(\theta) = \omega_0 [1+Ae^{-\theta^2/(2w^2)}]$ causes a constriction around $\theta = 0$. Here $\rho_0$ is the radius of the ring, $\omega_0$ the axial trapping frequency, $\lambda$ the ratio of transversal and axial trapping frequencies, and $A$ and $w$ are dimensionless numbers that characterize the strength and width of the link, respectively. To study a rotating link we consider a frame that moves along with it, adding to the right-hand side of Eq.~(\ref{egpe}) the term $-\Omega(t) L_z \psi$, where $\Omega(t)$ is the time-dependent rotating frequency, and $L_z = i\hbar(y\partial_x - x\partial_y)$. To obtain the ground state at some imposed angular momentum $L_0$ we minimize the quantity $\Tilde{E} = E + C(L-L_0)^2$, where $E$ is the energy corresponding to Eq.~(\ref{egpe}), $L = \int d \mathbf{r} \psi^\ast L_z \psi$, and $C > 0$ a number that, when sufficiently large, ensures that the energetic minimum occurs at $L \approx L_0$ \cite{Komineas2005}. We solve the extended Gross-Pitaevskii equation in imaginary and real-time by using the split-step Fourier method.
\section{Results}
\begin{figure}[ht]
\centering
\includegraphics[width = \columnwidth]{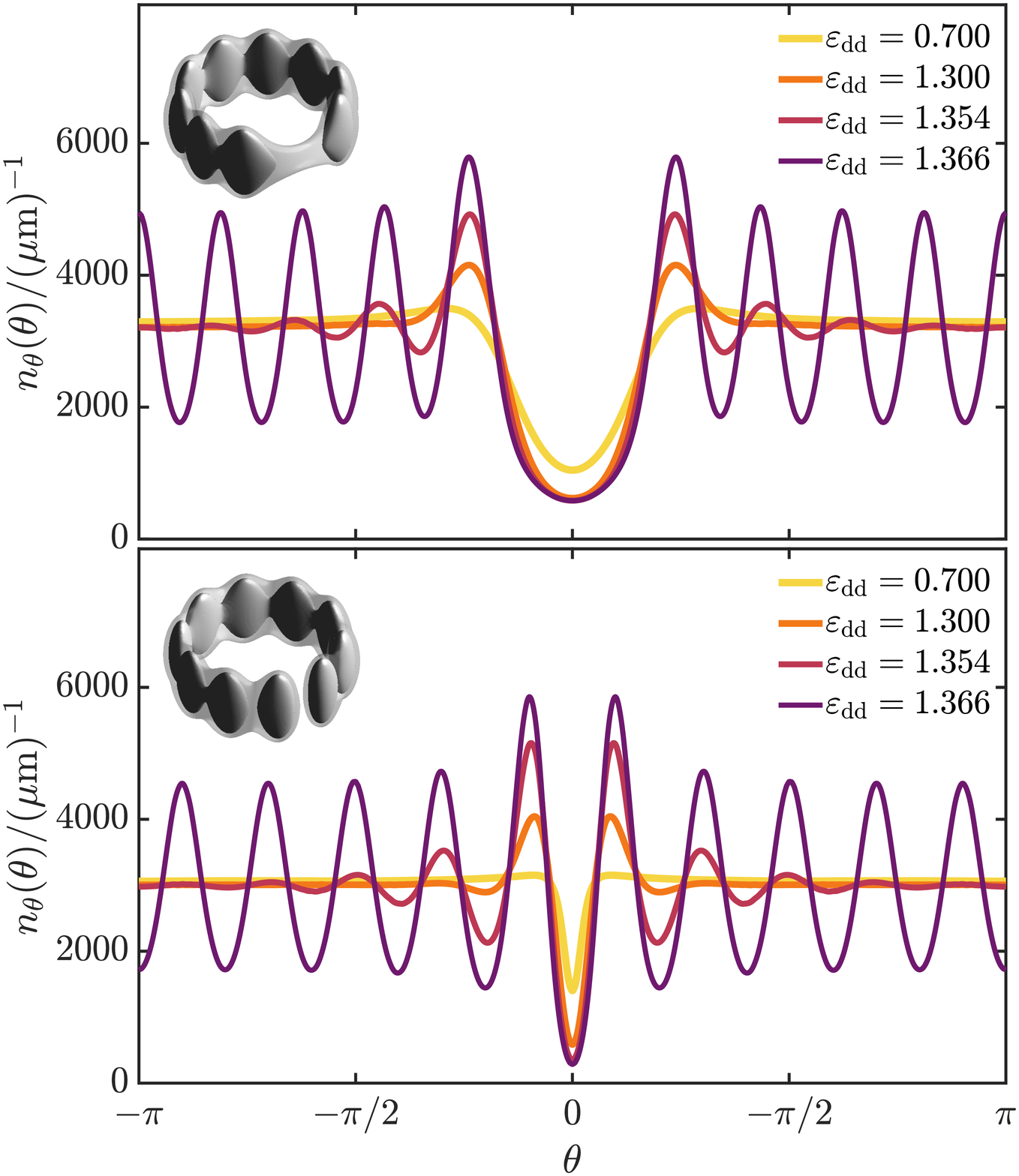}
\caption{({\it Color online}) Densities $n_\theta(\theta) = \iint d\rho dz |\psi|^2$ as a function of the azimuthal angle $\theta$ when $\varepsilon_\mathrm{dd} = 0.7, 1.3, 1.354, 1.366$ as indicated by the legend. The top panel shows densities for the wide link ($w=0.3$) and the bottom panel for the thin link ($w=0.03$). The insets in the top left corner of each panel show density isosurfaces for the case $\varepsilon_\mathrm{dd} = 1.366$, where the transparent and solid isosurfaces are taken at density values $n=7.5\times 10^{13}$ cm$^{-3}$ and $n=2\times 10^{14}$ cm$^{-3}$, respectively.}
\label{fig:angdens}
\end{figure}

\begin{figure}[ht]
\centering
\includegraphics[width = \columnwidth]{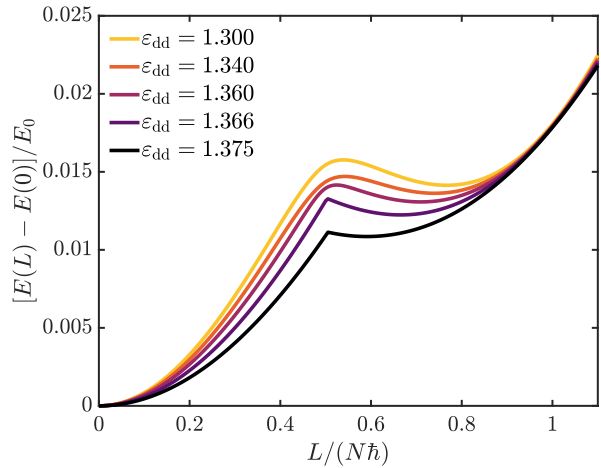}
\caption{({\it Color online}) Ground-state energy as a function of angular momentum for the wide ($w=0.3$) link when $\varepsilon_\mathrm{dd} = 1.3, 1.34, 1.36, 1.366, 1.375$, as indicated by the legend. The energy unit is $E_0 = \hbar^2/(m\xi^2)$, where $\xi = 1$ $\mu\mathrm{m}$.}
\label{fig:EL}
\end{figure}

\begin{figure}[ht]
\centering
\includegraphics[width = \columnwidth]{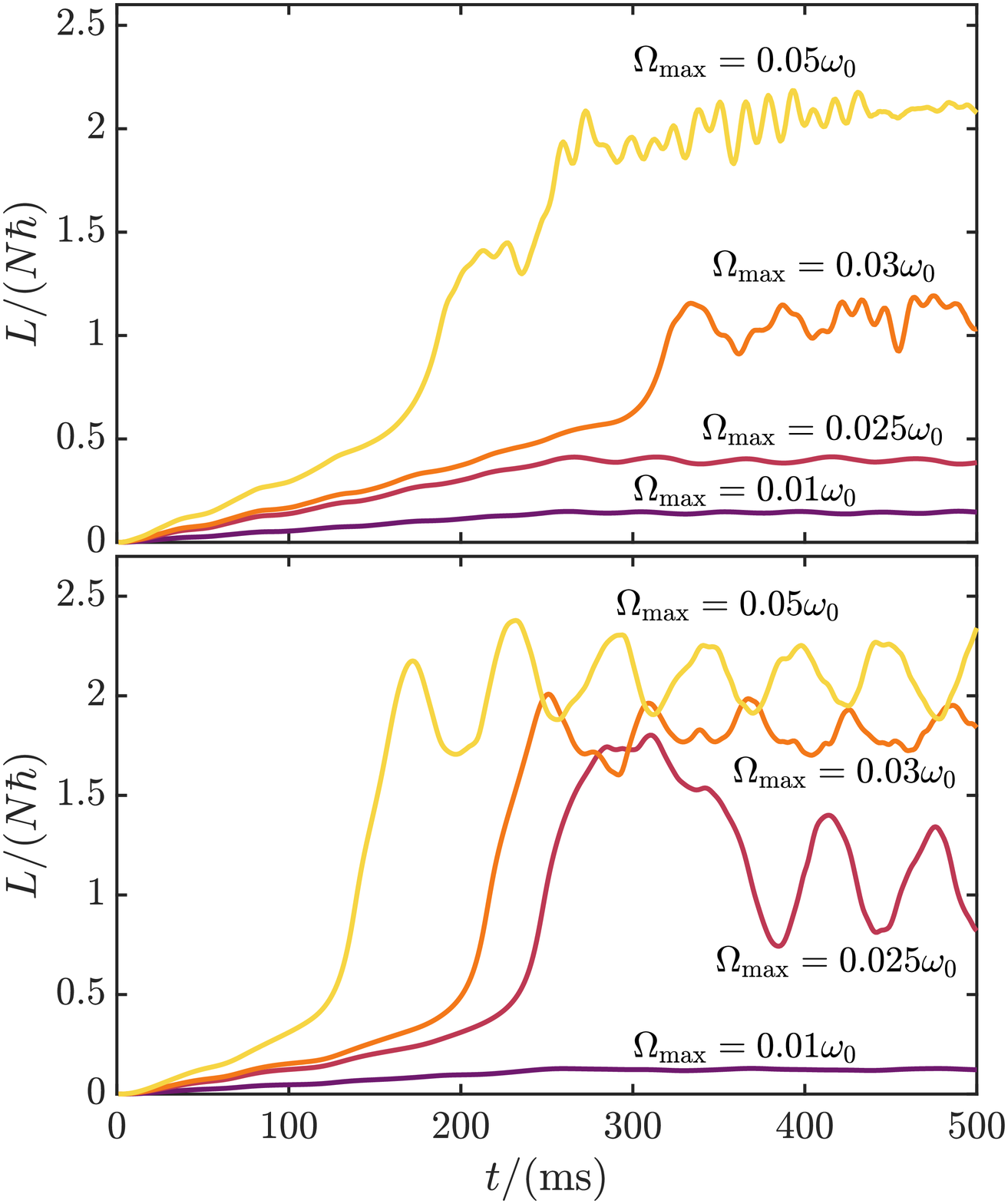}
\caption{({\it Color online}) Angular momentum as a function of time when the link is set to rotate for $\varepsilon_\mathrm{dd} = 1.366$ for the wide link ({\textit{top panel}}) and the thin link ({\textit{bottom panel}}). The rotation frequency is ramped linearly from $\Omega=0$ to $\Omega=\Omega_\mathrm{max}$ over 250 ms and then kept at this maximal value for another 250 ms.}
\label{fig:stir}
\end{figure}

\begin{figure}[t]
\centering
\includegraphics[width = \columnwidth]{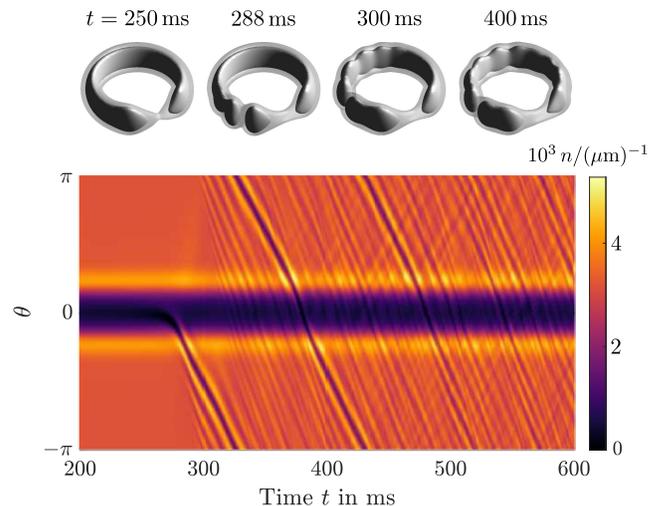}
\caption{({\it Color online}) ({\textit{top panel}}) Density distribution snapshots at given times with isosurfaces taken at $n=7.5\times10^{13}\,\mathrm{cm}^{-3}$ and $n=2.0\times10^{14}\,\mathrm{cm}^{-3}$. Note the varying amount of droplets throughout. ({\textit{bottom panel}}) Time evolution of the integrated angular density $n_\theta(\theta) = \iint d\rho\, dz|\psi|^2$ for $\varepsilon_{\mathrm{dd}}=1.3$.}
\label{fig:roton_sf}
\end{figure}

\begin{figure}[t]
\centering
\includegraphics[width = \columnwidth]{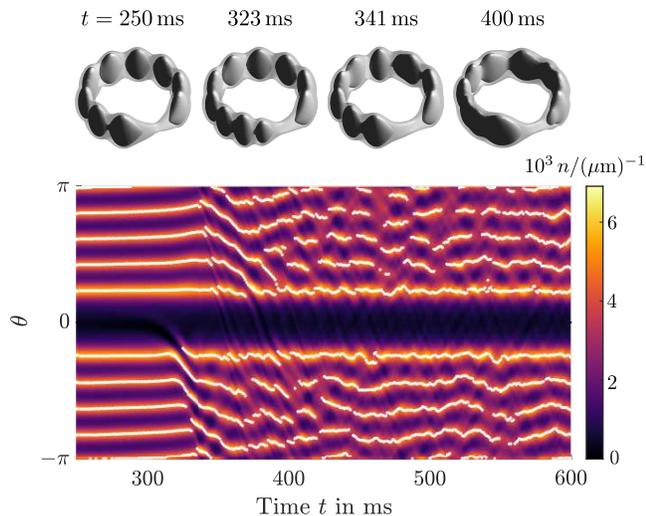}
\caption{({\it Color online}) ({\textit{top panel}}) Density distribution snapshots at given times with isosurfaces taken at $n=7.5\times10^{13}\,\mathrm{cm}^{-3}$ and $n=2.0\times10^{14}\,\mathrm{cm}^{-3}$. Note the varying amount of droplets throughout. ({\textit{bottom panel}}) Time evolution of the integrated angular density $n_\theta(\theta) = \iint d\rho\, dz|\psi|^2$ for $\varepsilon_{\mathrm{dd}}=1.366$. White dots mark the positions of each local maxima with $n_\theta(\theta)>4000\,(\mu\mathrm{m})^{-1}$.}
\label{fig:roton_ss}
\end{figure}

To exemplify our findings we consider a system consisting of $^{164}$Dy atoms and fix the parameters $N = 1\times 10^5$, $\rho_0 = 5$~$\mu$m, $\omega_0 = 100$~Hz, and $\lambda = 1.1$. The amplitude of the weak link is set to $A = 0.5$ with a width equal to $w = 0.3$ for a wide link or $w=0.03$ for a thin link.\newline
For these parameters, we find that in the absence of a weak link the ground state in the supersolid phase contains ten localized high-density sites. Adding a weak link this amount reduces by one for the wide link and remains unchanged for the thin link (see Fig.~\ref{fig:angdens}) where the integrated density along the azimuthal angle $n_\theta(\theta) = \iint d\rho dz |\psi|^2$ is displayed for some values of $\varepsilon_\mathrm{dd}$. The emergence of the supersolid phase in the presence of a weak link is perhaps not as evident as for the system without, since in the latter the supersolid transition in the ring occurs precisely when azimuthal symmetry breaks. This spontaneous symmetry breaking can not be taken as the transition point for the weak-link system since the trap itself breaks azimuthal symmetry.\newline
We can see that when the dipolar interaction is weak compared to the contact interaction the density is nearly homogeneous away from the link. This behavior changes with increased relative interaction strength, and the atoms start to cluster at the edges of the weak link, as shown here for $\varepsilon_\mathrm{dd}=1.3$. This can be understood since the diminished density in the link results in a decreased side-by-side repulsion in its vicinity. For stronger interactions, the density acquires an oscillatory behavior where the amplitude diminishes away from the link. For even larger values of $\varepsilon_\mathrm{dd}$, the overall behavior stays the same as the oscillation amplitude increases and clear regular localizations in the density appear.\newline
To gain some insight regarding the rotational properties of the weak-link system, we consider the ground-state energy as a function of the angular momentum for different values of $\varepsilon_\mathrm{dd}$, see Fig.~\ref{fig:EL} for results with a wide link (we find the same qualitative behavior for $w=0.3$ and $w=0.03$ and choose to display results only for the former here). In the superfluid phase, the energy forms dampened intersecting parabolas, similar to a mixed Bose-Bose system in the bubble phase~\cite{Sturmer2022}. Such dampened behavior builds the basis of the neutral atomic qubit occurring for repulsive Bose gases in the superfluid phase~\cite{ref:Anderson2003,ref:Solenov2010a,ref:Solenov2010b,ref:Ryu2013,Cominotti2014,ref:Aghamalyan2015,ref:Amico2021,ref:Amico2022}. The dampened behavior is in contrast to a superfluid in an azimuthally symmetric torus, where $E(L)$ is concave with a V-shaped minimum at $L/(N\hbar) = 1$ \cite{ref:Bloch1973}. As we increase the relative interaction strength and approach the supersolid phase, the intersection of the parabolas sharpens and moves towards $L/(N\hbar)\sim 0.5$, and eventually, the energetic behavior becomes that of two intersecting parabolas. This is similar to the toroidal dipolar supersolid without a link \cite{ref:NilssonTengstrand2021}, and has the interpretation of rigid motion of the localized high-density regions together with counterflow in the low-density regions for angular momenta smaller than $L/(N\hbar) = 0.5$, and rigid motion plus a singly quantized vortex nucleated in the center for angular momenta larger than $L/(N\hbar) = 0.5$ (and smaller than $L/(N\hbar) = 1.5$, beyond which multiply-quantized vortices become energetically favorable). Since local energetic minima exist at finite $L>0$ for small enough $\varepsilon_\mathrm{dd}$, the supersolid may consequently support persistent currents when a weak link is present.\newline
An interesting question is whether it is possible to introduce vorticity to a supersolid through the rotation of a weak link. To this end, we consider a protocol where the link's rotation frequency $\Omega$ is ramped linearly from $\Omega=0$ to $\Omega = \Omega_\mathrm{max}$ over 250 ms and then kept at this maximal value for another 250 ms. Fig.~\ref{fig:stir} shows the angular momentum as a function of time $L(t)$ for both links for four different values of $\Omega_\mathrm{max}$ \footnote{See the Supplemental Material for movies of the full time-evolution of the corresponding densities and phases}. Before the sharp jumps in angular momentum, all systems oscillate regularly in $L/(N\hbar)$. This oscillation originates from the excitation of a mode corresponding to the swaying back and forth of the localized high-density regions of the supersolid. When the maximal rotation frequency is sufficiently high, $L(t)$ is characterized by a quick increase in the angular momentum, which corresponds to the entering of vortices into the condensate. We see that the behavior for the different links is quite different and that the critical rotation frequencies also differ. To estimate this frequency (where an $s$-times quantized vortex is energetically favorable) it is possible to use the analytical expression derived in Ref.~\cite{ref:NilssonTengstrand2021} for a torus without a weak link
\begin{equation}
\Omega_\mathrm{crit} = \frac{(2s-1)\hbar}{2m\rho_0^2},
\end{equation}
which was derived for a system where the radius of the torus is much larger than the length scale set by the trapping frequency, i.e. $\rho_0 \gg \sqrt{\hbar/(m\omega_0)}$. For the parameters in this study this translates to $\Omega_{\mathrm{crit}} / \omega_0 \approx 0.012$ for $s=1$, $\Omega_{\mathrm{crit}} / \omega_0 \approx 0.037$ for $s=2$, and $\Omega_{\mathrm{crit}} / \omega_0 \approx 0.062$ for $s=3$. Although these values serve as decent estimates, an inspection of the results in Fig.~\ref{fig:stir} shows that these estimates are not exact. In particular, for the thin link when $\Omega_\mathrm{max} = 0.025$, two vortices enter in quick succession around $t = 250$ ms followed by the almost immediate exit of one of these, leaving only a single vortex remaining in the system. Following the quick increase in angular momentum associated with vorticity, $L(t)$ becomes strongly oscillatory. As the vortices enter, a solitonic excitation appears that moves more slowly than the link. This results in an interference between the solitonic excitation and the localized structure of the supersolid, and gives rise to additional oscillations on top of the already excited mode. \newline
To understand this mechanism, let us investigate the entry of solitonic excitations due to the phase slip associated with the stirring protocol. In such a case, solitonic excitations provide a pathway for non-quantized currents. Firstly, we recover the effects known for a Bose-gas in the phonon regime (either purely contact interacting or at low relative dipolar interaction strengths, here $\varepsilon_{\mathrm{dd}}=0.7$) in a torus with a weak link~\cite{ref:Munoz2015}. In this case, a vortex-antivortex pair creates the counterflow in the region of the weak link, similar to the Meissner effect~\cite{ref:Woo2012,ref:Dubessy2012}. The resulting phase jump increases with rotational speed until the flow velocity inside the weak link is equal to the local speed of sound. At this point a phase slip occurs, creating a solitonic vortex that moves along the torus~\cite{ref:Abad2011,ref:Yakimenko2015}. This solitonic excitation decays by creating phononic excitations itself, moving closer to the outer edge of the superfluid in the process.\newline
For a system close to the roton-instability (here for $\varepsilon_{\mathrm{dd}}=1.3$, see Fig.~\ref{fig:roton_sf}) this process is equal up to the point of the emission of solitonic excitations. Here we find a dark solitary wave moving through the condensate instead of a solitonic excitation. Similarly to the above, the solitary wave creates quasi-particle excitations which move with the local speed of sound. However, the excited quasi-particles are rotons. This triggers the roton instability, such that one observes a periodic density modulation preceding the dark solitary wave with the local speed of sound. The density modulation itself disappears and reemerges periodically as the solitary wave and rotonic excitation move through the system.\newline
Equipped with these insights, let us now turn to the supersolid phase (see Fig.~\ref{fig:roton_ss}). As the phase slip occurs, similar to the superfluid case in the roton regime, the dark solitary wave emerges. However, here the solitary wave becomes untraceable as it encounters the first droplet. As before, the solitary wave excites rotons, which move at the local speed of sound. Additionally, the density modulation associated with the solitary wave creates a new droplet between the solitary wave and the weak link. After the solitary wave decays, the initial kick and propagation of the excited rotons determine the dynamics and propagation of the system. These rotons interact with the existing droplets, leading to a continuous melting and recrystallization of droplets. If two or more droplets move too close to each other, they may melt into a large, lower-density area (see the first two droplets below the weak link at around $400$ ms in Fig.~\ref{fig:roton_ss}). This melting is similar to the mean displacement argument in Lindemann's melting criterion \cite{ref:Lindemann1910}, where one, based on a system's temperature, estimates the solid-liquid phase transition by calculating the mean displacement necessary for two crystal sites to touch. On the other hand, if two individual droplets move too far apart, a new droplet forms in between them (see the first two droplets below the weak link at around $550$ ms in Fig.~\ref{fig:roton_ss}). The propagating roton-excitation further mitigate this behavior when interacting with the already existing structure.
All three cases outlined above for the wide link are equivalent to the tight link. In the supersolid case, however, the dark solitary wave dissipates much faster.\newline
\begin{figure}[t]
\centering
\includegraphics[width = \columnwidth]{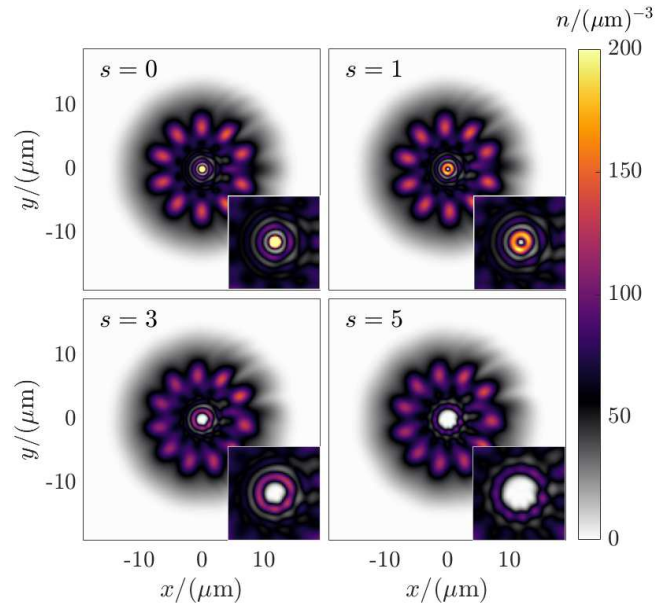}
\caption{({\it Color online}) Densities in the $z=0$ plane 5 ms after the radial trap has been turned off for the thin link ($w=0.03$) for different values of the vortex quantization $s=0,1,3,5$, as indicated by the figure. The relative interaction strength is $\varepsilon_\mathrm{dd} = 1.366$ and the initial systems prior to release are generated as ground states in the rotating frame at rotation frequencies $\Omega/\omega_0 = 0, 0.02, 0.07, 0.11$, respectively. Note that the color scale is saturated and that the peak density at the center of the top left panel is $n \approx 570$ $(\mu \mathrm{m})^{-3}$. The insets in the bottom right corner of each panel show a zoom-in of the central region.}
\label{fig:release}
\end{figure}
Finally, we look at the release of a supersolid from the weak-link trap for configurations with different amounts of initial vorticity, created by obtaining ground states in a rotating frame. Fig.~\ref{fig:release} shows densities in the $z=0$ plane 5 ms after the radial trap has been turned off (thus also removing the link). When there is no initial vortex ($s=0$), a high-density region forms at the center due to interference of the dispersing condensate (see the top left panel in Fig.~\ref{fig:release}). Such a central density peak is prevented when $s\neq0$ due to the existence of a vortex core there, and the size of this core increases with increasing $s$, similarly to what occurs for a regular BEC made out of $^{23}$Na atoms \cite{Murray2013}. The coexistence of superfluidity and a crystalline structure is thus clearly established for the vortical supersolid in this type of time-of-flight setup, exhibiting both a central hole and a discrete localized structure.
\section{Conclusions and Outlook}
In conclusion, we have studied dipolar supersolids in ring-shaped traps with a weak link. Through a study of the energetic behavior, we found that persistent currents may be supported in these systems. Simulating the real-time dynamics of a rotating link, we saw that the entry of vortices resulted in the creation of solitonic excitations traveling along the ring, interacting with the supersolid structure. When released from the radial trap, we observed a characteristic density pattern with a discrete structure together with a central hole, providing a useful experimental signature for the coexistence of superfluidity and spontaneous density modulation. Superfluid Bose-Einstein condensates made out of $^{23}$Na in ring-shaped traps with a weak link have been realized experimentally \cite{Wright2013, Eckel2014}, and the parameters used throughout this paper are of similar orders of magnitude as in these experiments. Unlike the pure superfluid, however, the high-density regions in the supersolid phase set a boundary on the lifetimes due to the increased impact of three-body recombination losses. Finally, it would be interesting to study the behavior of an asymmetric binary Bose mixture in the droplet-superfluid phase in the presence of a weak link. These systems are analogous to the dipolar supersolids studied here in many ways \cite{ref:Petrov2015, NilssonTengstrand2022}, and may help pave the way in increasing our understanding of beyond-mean-field physics in the context of atomtronics.
 
\begin{acknowledgments}
This work was financially supported by the Knut and Alice Wallenberg Foundation, the Swedish Research Council, and NanoLund. We are grateful for many fruitful discussions with K. Mukherjee, T. Arnone Cardinale, T. Bland, as well as A. Recati and S. Stringari at the initial stage of the project.
\end{acknowledgments}

\bibliography{Link.bib}

\end{document}